\documentclass[10pt,twocolumn,letterpaper]{article}

\usepackage{iccv}
\usepackage{times}
\usepackage{epsfig}
\usepackage{graphicx}
\usepackage{amsmath}
\usepackage{amssymb}

\usepackage{xcolor}
\usepackage{pstricks}
\usepackage{graphicx}
\usepackage{caption}
\usepackage{subcaption}
\usepackage[normalem]{ulem}
\usepackage{xspace}
\usepackage{cancel}
\usepackage{booktabs}
\usepackage{siunitx}
\usepackage[inline]{enumitem}
\usepackage{pifont}
\usepackage{graphicx}
\usepackage{tabularx}
\usepackage{animate}

\usepackage[pagebackref=true,breaklinks=true,letterpaper=true,colorlinks,bookmarks=false]{hyperref}

\iccvfinalcopy 
\graphicspath{ {./figures/} {./figures/placeholders/} }

\def\iccvPaperID{10117} 

\ificcvfinal\fi

\newcommand{\fref}[1]{Fig.~\ref{#1}}
\newcommand{\tref}[1]{Table~\ref{#1}}
\newcommand{\eref}[1]{Eq.~\ref{#1}}

\newcommand{\cref}[1]{Chapter~\ref{#1}}
\newcommand{\sref}[1]{Sec.~\ref{#1}}

\definecolor{red}{rgb}{0.8,0,0}
\definecolor{purered}{rgb}{1,0,0}
\definecolor{darkred}{rgb}{0.6,0,0}
\definecolor{green}{rgb}{0.0,0.5,0}
\definecolor{blue}{rgb}{0,0,0.75}
\definecolor{lightblue}{rgb}{0.3,0.3,0.75}
\definecolor{darkblue}{rgb}{0,0,0.55}
\definecolor{orange}{rgb}{0.9,0.3,0.1}
\definecolor{purple}{rgb}{0.6,0.0,0.6}
\definecolor{cyan}{rgb}{0.0,0.7,0.7}
\definecolor{darkgray}{rgb}{0.4,0.4,0.4}
\definecolor{bronze}{rgb}{0.8, 0.5, 0.2}
\definecolor{dorange}{rgb}{0.75, 0.4, 0.0}

\newcommand{\iccvfinal}[1]{#1}
\newcommand{\iccv}[1]{#1}

\newcommand{\shortcite}[1]{\cite{#1}}

\newcommand{\diff}{\mathrm{d}}

\newcommand{\eqbreak}{\nonumber \\}
\newcommand{\x}{\ensuremath{\mathbf{x}}\xspace}

\newcommand{\xs}{\ensuremath{\x_{s}}\xspace}

\newcommand{\stransportImage}{\mathbf{i}}
\newcommand{\stransportMatrix}{\mathbf{T}}
\newcommand{\stransportMatrixTransient}{\mathbf{H}}
\newcommand{\stransportSources}{\mathbf{p}}

\newcommand{\svtransportMatrix}{\stransportMatrix}

\newcommand{\sdefocusMatrix}{\mathbf{D}}

\newcommand{\norm}[1]{\left\lvert#1\right\rvert}

\newcommand{\pftime}{t}
\newcommand{\planeC}{S}
\newcommand{\planeP}{L}
\newcommand{\xp}{\mathbf{x}_l}
\newcommand{\vvec}{\vec{\mathbf{v}}}
\newcommand{\dxp}{\diff \xp}
\newcommand{\xc}{\mathbf{x}_s}
\newcommand{\dxc}{\diff \xc}

\newcommand{\xv}{\mathbf{x}_v}
\newcommand{\xa}{\mathbf{x}_a}
\newcommand{\xb}{\mathbf{x}_b}
\newcommand{\nonEmpty}{\Gamma}
\newcommand{\maskNonEmpty}{\textbf{m}_{\nonEmpty}}
\newcommand{\phasor}{\mathcal{P}}
\newcommand{\gating}{{G}}
\newcommand{\dist}{{D}}
\newcommand{\distf}[1]{\dist\left(#1\right)}
\newcommand{\gatingf}[1]{\gating\left(#1\right)}
\newcommand{\phasorf}[1]{\phasor\left(#1\right)}
\newcommand{\phasorwf}[1]{\phasor_{\pfFreq}\!\left(#1\right)}
\newcommand{\phasorxt}{\phasorf{\x, \pftime}}
\newcommand{\phasorwxt}{\phasorwf{\x, \pftime}}
\newcommand{\phasorxpt}{\phasorf{\xp, \pftime}}
\newcommand{\phasorwxpt}{\phasorwf{\xp, \pftime}}
\newcommand{\phasorxct}{\phasorf{\xc, \pftime}}
\newcommand{\phasorwxct}{\phasorwf{\xc, \pftime}}
\newcommand{\pfImagingModel}{\Phi}
\newcommand{\ROI}{V}
\newcommand{\pfImage}{I}
\newcommand{\pfImageof}[2]{\pfImage_{#1}\left(#2\right)}
\newcommand{\pfImpulse}{\stransportMatrixTransient}
\newcommand{\pfImpulseFun}{\pfImpulse\left(\xp,\xc,\pftime\right)}
\newcommand{\tof}{\text{\pftime}}

\newcommand{\pfThinLens}{\mathcal{L}}
\newcommand{\pfThinLensFun}[2]{\pfThinLens_{#1}\!\left(#2\right)}
\newcommand{\pfFreq}{\omega}

\newcommand{\pathseq}[1]{\left\langle#1\right\rangle}

\begin{document}

\title{Virtual light transport matrices for non-line-of-sight imaging}

\author{Julio Marco$^1$\quad
	Adrian Jarabo$^1$\quad 
	Ji Hyun Nam$^2$\quad
	Xiaochun Liu$^2$\quad \\
	Miguel Ángel Cosculluela$^1$\quad 
	Andreas Velten$^2$\quad 
	Diego Gutierrez$^1$\\ \\
	$^1$Universidad de Zaragoza, I3A \quad $^2$University of Wisconsin -- Madison
}

\maketitle
\ificcvfinal\thispagestyle{empty}\fi

\begin{abstract}
	\iccv{The light transport matrix (LTM) is an instrumental tool in line-of-sight (LOS) imaging, describing how light interacts with the scene and enabling applications such as relighting or separation of illumination components. 
We introduce a framework to estimate the LTM of non-line-of-sight (NLOS) scenarios, coupling recent virtual forward light propagation models for NLOS imaging with the LOS light transport equation. 
%
%
%
%
%
%
%
We design computational projector-camera setups, and use these virtual imaging systems to estimate the transport matrix of hidden scenes. We introduce the specific illumination functions to compute the different elements of the matrix, overcoming the challenging wide-aperture conditions of NLOS setups.  
Our NLOS light transport matrix allows us to (re)illuminate specific locations of a hidden scene, and separate direct, first-order indirect, and higher-order indirect illumination of complex cluttered hidden scenes, similar to existing LOS techniques.}
\end{abstract}

\begin{figure}
    \centering
    \includegraphics[width=0.87\linewidth]{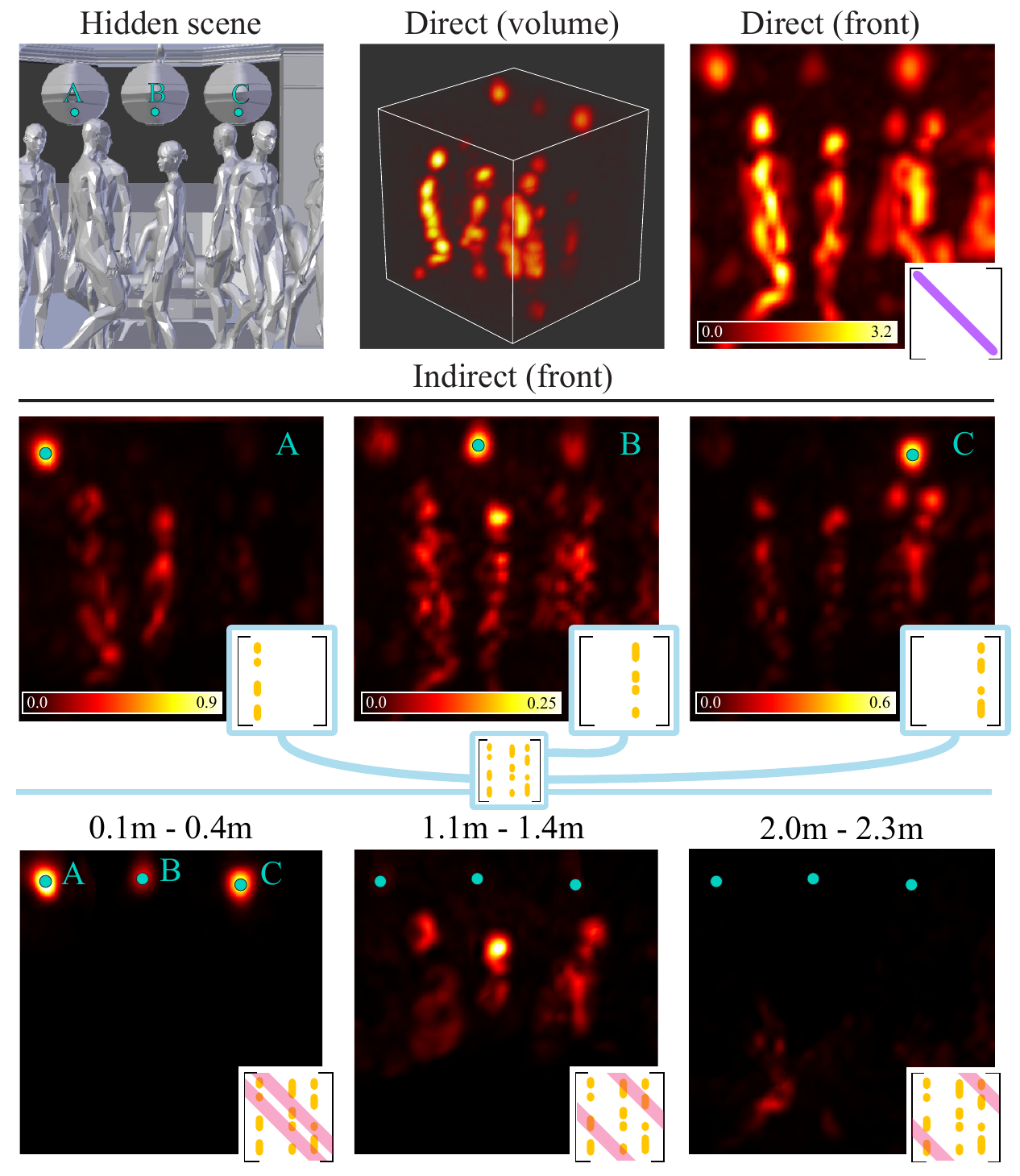}%
    \caption{\iccv{We introduce a framework to compute virtual light transport matrices (LTM) of NLOS scenes. 
Probing the virtual LTM of a hidden scene allows us to extract its direct illumination (first row), the indirect components when illuminating a single point in the scene (second row, each point corresponds to a column in the LTM), or decomposing near-, middle-, and far-field indirect components (third row) when illuminating specific points in the hidden scene. The insets show the probed elements from the virtual LTM for each image. }
 %
%
}

    \label{fig:teaser}
    \vspace{-1em}
\end{figure}

\vspace{-1em}
\section{Introduction}
\iccv{



The light transport matrix (LTM) \cite{Ng2003} is a fundamental tool to understand how a scene transports incident light, describing the linear response of the scene for a given illumination. It has become the cornerstone of many applications such as dual photography \cite{Sen2005Dual}, image-based relighting~\cite{o2010optical,Wang2009kernel}, acquisition of material properties~\cite{Debevec2000Acquiring,peers2006compact}, separation of global illumination components  \cite{Nayar2006,OToole2012}, or robust depth estimation \cite{otoole2014}. To acquire the LTM, a camera-projector setup is used, capturing the scene under varying coded illumination conditions. 

On the other hand, recent works on non-line-of-sight (NLOS) imaging have shown the potential to significantly advance many fields including medical imaging, autonomous driving, rescue operations, or defense and security, to name a few. The key idea is to leverage the scattered radiance on a secondary relay surface to infer information about the hidden scene~\cite{Velten2012nc,heide2019non,maeda2019recent}. However, NLOS imaging is still on its infancy, and many well-established capabilities of traditional line-of-sight (LOS) imaging modalities cannot yet be applied in NLOS conditions.


In this work we take steps towards bridging this gap between LOS and NLOS imaging modalities, introducing a computational framework to obtain the LTM of a hidden scene. In particular, we build on the recent wave-based \emph{phasor fields} framework~\cite{Liu2019phasor}, which  poses NLOS as a \emph{forward} wave transport problem, creating virtual (computational) light sources and cameras at the visible relay surface from time-resolved measurements of the hidden scene.

While this approach would \textit{in principle} allow turning NLOS imaging into a \textit{virtual LOS} problem, working in the NLOS domain introduces two main challenges which are not present in LOS settings: 
%
\iccvfinal{1) the large baseline of capture setups on the visible surfaces results in a very large virtual aperture, creating a very shallow depth-of-field and therefore significant out-of-focus contribution; and 2) the resolution of the virtual projector-camera is limited, with potentially significant cross-talk between neighbor (virtual) pixels.}
As a consequence, directly applying LOS techniques to capture the LTM of hidden scenes would lead to suboptimal results.

We develop computational (virtual) illumination and imaging functions, leveraging the fact that undesired contributions of light transport in the hidden scene are coded in its NLOS virtual LTM. 
Then, inspired by existing works in light transport analysis in LOS settings~\cite{Nayar2006,OToole2012}, we exploit the LTM and demonstrate illumination decomposition in NLOS scenes (see \fref{fig:teaser}). \iccvfinal{In particular, we probe} different elements of the matrix, including direct and indirect illumination, as well as near-, middle-, and far-field indirect light decomposition. This has potential applications for relighting, material analysis, improved geometry reconstruction, or scene understanding in general beyond the third bounce. In summary, our \textbf{contributions} are:
\begin{itemize}
	\item We introduce a framework to obtain the light transport matrix (LTM) of a hidden scene, helping bridge the gap between LOS and NLOS imaging. 
	\item We develop the computational methods to obtain the necessary virtual illumination and imaging functions, dealing with the main challenges on the NLOS imaging modality. 
	\item We demonstrate how our formulation allows to probe the virtual LTM, allowing to separate NLOS direct and indirect illumination components in hidden scenes. 
\end{itemize}
}



\section{Related work}
\paragraph{\iccv{Light transport acquisition and analysis.}} 
Several works have focused on acquiring the linear light transport response from a illuminated scene  (i.e., the \emph{light transport matrix}) by means of exhaustive capture~\cite{Debevec2000Acquiring}, low-rank matrix approximations~\cite{Wang2009kernel} and homogeneous factorization~\cite{OToole2015}, optical computing~\cite{o2010optical}, compressed sensing~\cite{Peers2009Compressive,Sen2009}, or machine learning~\cite{xu2018deep}. \iccv{In our work we propose a framework to compute virtual transport matrices of NLOS scenes using virtual camera/projector systems, which could be potentially combined with any of these approaches for improved efficiency.} 

In terms of light transport analysis, Sen et al.~\shortcite{Sen2005Dual} leveraged the Helmholtz reciprocity to create a dual camera-projector system allowing to take photographs with a single pixel camera. Nayar et al.~\shortcite{Nayar2006} exploited the structure of the linear light transport operator to disambiguate between direct and indirect illumination. Mukaigawa et al.~\shortcite{Mukaigawa2010analysis} generalized this work to scattering media. Later, Nayar et al.'s approach was improved by O'Toole and Kutulakos~\shortcite{OToole2012}, who introduced advanced probing codes in the illumination, allowing to disambiguate between high- and low-frequency global illumination. Follow-up work~\cite{otoole2014} included the temporal domain, allowing high-quality depth reconstruction using time-of-flight cameras. \iccv{Our paper builds on these works, but shifts its domain from visible scenes to the regime of NLOS scenes, allowing to obtain virtual light transport matrices of occluded scenes}.
\paragraph{Impulse NLOS imaging.}
\iccv{Transient imaging methods leverage information encoded in time-resolved light transport using ultra-fast illumination capture systems \cite{Velten2013,Heide2013,Gariepy2015} in applications such as light-in-flight videos~\cite{Velten2013,Heide2013,OToole2017}, bare-sensor imaging~\cite{Wu2012eccv}, or seeing through turbid media~\cite{Heide2014, Wu2018adaptive}. In the following we discuss methods related to NLOS imaging based on transient light transport, and refer to the reader to \cite{Jarabo2017transient} for a broader overview.}

This line of work follows the approach proposed by Kirmani et al.~\shortcite{Kirmani2009ICCV} and demonstrated experimentally by Velten and colleagues~\shortcite{Velten2012nc}: very short laser pulses illuminate a visible surface facing the hidden scene, then the scattered light reflected back onto such visible surface is captured. The hidden scene is then reconstructed using backprojection algorithms and heuristic filters~\cite{Velten2012nc,Laurenzis2014area,Arellano2017NLOS,Chen2020learned}, or inverting simplified light transport models~\cite{o2018confocal,Ahn2019,Pediredla2019SNLOS,Xin2019theory,heide2019non,Young2020NLOS,iseringhausen2020non}. 
These methods have usually been demonstrated in simple isolated scenes with little indirect illumination from interreflections. In contrast, recent wave-based methods for NLOS imaging~\cite{Liu2019phasor,Lindell2019wave,Liu2020phasor} removed these limitations \iccv{by posing NLOS imaging as \emph{forward} virtual propagation models. 
We leverage existing formulations of forward propagation~\cite{Liu2019phasor} to compute virtual LTMs of hidden scenes using virtual projector/camera pairs. Different from (and complementary to) classic NLOS geometry reconstruction approaches, our work focuses on analyzing light transport, paving the way for relighting and separation of illumination components in hidden scenes. Wu et al.~\shortcite{Wu2014} and Gupta et al.~\cite{Gupta2015phasor} used transient imaging to disambiguate between direct and global light transport in a LOS scene; in contrast, we demonstrate direct-indirect separation in \emph{non-line-of-sight} scenes by estimating and probing their light transport matrices.}


\section{Background}
\label{sec:background}
%
In the following we summarize the key aspects of the light transport matrix and the forward NLOS formulation of phasor fields. \tref{tab:notation} defines a list of the most common symbols and expressions used throughout the paper.
\begin{table}\small 
    \centering
    \caption{Notation used throughout the paper.}
    \begin{tabularx}{\columnwidth}{l X}    
	\toprule  
	$V$ 											& Voxelized space of the hidden scene. \\
	$K_{\textbf{v}}$ 								& Number of voxels discretizing $V$. \\
	$\nonEmpty\subset V$ 							& Subspace of $V$ of non-empty regions. \\
	$\xa, \xb, \xv\! \in\! V$								& Points in the hidden scene. \\
	$\planeP,\planeC$ 								& Laser and SPAD capture planes.\\
	$K_{\stransportSources}, K_{\stransportImage}$ & Number of measurements in $\planeP$ and $\planeC$. \\
	$\xp\!\in\!\planeP, \xc\!\in\!\planeC$					& Points at planes $\planeP$ and $\planeC$, respectively.\\
	$\pathseq{\x_0,...,\x_n}$ 						& A $(n+1)$-vertex light path from $\x_0$ to $\x_n$.\\
	\midrule
	$\phasorxt$ 										& Broadband complex phasor at point $\x$. \\
	$\phasorwxt$ 									& Monochromatic phasor at $\x$ with frequency $\pfFreq$. \\
	$\pfThinLensFun{\pfFreq}{\x_a,\x_b,\pftime}$ 	& Monochromatic thin-lens phasor propagator from $\x_a$ to $\x_b$ with frequency $\pfFreq$.\\
$\gatingf{t^\prime, t}$ 							&  Virtual temporal gating function centered at time $t'$. \\
	\midrule
	$\stransportMatrix(\xa,\xb)$					& Virtual transport matrix at illuminated and imaged locations $\xa$ and $\xb$. \\
	$ \mathbf{M}_{i}(\xa,\xb)$ 					& Matricial binary mask removing the contribution from $\xa$ to $\xb$\\
	$\pfImpulseFun$									& Time-resolved impulse response function at laser and SPAD locations $\xp$ and $\xc$. \\	
	\bottomrule
    \end{tabularx}
    \label{tab:notation}
\end{table}

\subsection{Transport matrix formulation}
\label{sec:lighttransport_background}
Light transport is a linear process that can be encoded in a transport matrix~\cite{Ng2003} relating the response of the scene captured by the sensor $\stransportImage$ with the illumination $\stransportSources$ as
\begin{align}
\stransportImage = \stransportMatrix \stransportSources, 
\label{eq:steadytransportmatrix}
\end{align}
where $\stransportSources$ is a $K_{\stransportSources}$-sized column vector representing the light sources, $\stransportImage$ is a $K_{\stransportImage}$-sized column vector representing the observations, and $\stransportMatrix$ is the transport matrix of size $K_{\stransportSources} \times K_{\stransportImage}$ modeling the linear response of the scene. Intuitively, each row of $\stransportMatrix$ allows us to reconstruct the captured scene as illuminated by a single light source $j$, represented as the $j$-{th} component of vector $\stransportSources$.
%
%
%
Under controlled illumination, the transport matrix $\stransportMatrix$ of a scene can be trivially computed by independent activation of all light sources $\stransportSources$ in the scene. Measuring the image response of each activated light source $\stransportSources_j$ allows us to fill the $j$-th row of $\stransportMatrix$. More efficient methods have been developed for reconstructing $\stransportMatrix$ from a sparse set of measurements \cite{Wang2009kernel, Peers2009Compressive,o2010optical,OToole2015}.

%

%

%

%

%
Probing the LTM~\cite{Nayar2006,OToole2012} allows to separate illumination components at fast frame rates using aligned arrays of emitters (projectors) and sensors (camera pixels), where both $\stransportSources$ and $\stransportImage$ have the same size $K_{\stransportSources} \equiv K_{\stransportImage}$. The diagonal of the resulting square matrix $\stransportMatrix$ represents a good approximation of direct illumination, since every image pixel is lit only by its co-located projector pixel. In contrast, non-diagonal elements of the transport matrix $\stransportMatrix$ encode the multiply scattered indirect light, where a column would represent the indirect contribution of a single emitter over the entire scene. 
In our work, we show how to compute the LTM on a hidden scene, by computationally creating a virtual projector-camera pair from a set of NLOS measurements. 

\subsection{Phasor field NLOS imaging}
\label{sec:phasor_fields_background}
\begin{figure}
	\centering
	\def\svgwidth{\linewidth}
	\fontsize{8pt}{10pt}\selectfont
	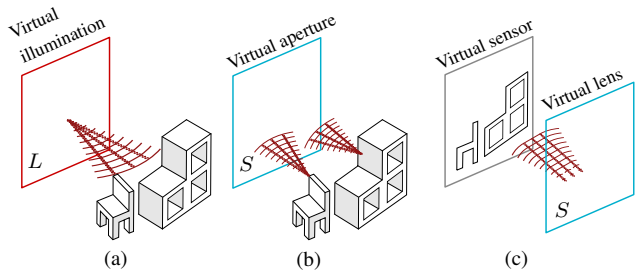%
	\caption{
The phasor field method leverages transient measurements on a visible wall to propagate virtual light wave fields (the phasor field) towards a hidden scene (a), then image its response (b) from a virtual line-of-sight camera perspective (c). Figure adapted from \cite{Liu2019phasor}.}
	\label{fig:phasor_fields_scheme}
\end{figure}
Light transport as described by \eref{eq:steadytransportmatrix} assumes steady state. By adding the temporal domain $t$ to the transport matrix,  thus taking into account propagation and scattering light transport delays, $\stransportMatrix$ becomes the linear \emph{impulse} response of the scene $\stransportMatrixTransient (t)$, and \eref{eq:steadytransportmatrix} becomes \cite{otoole2014}
\begin{align}
\stransportImage(t) & = \int_{-\infty}^\infty \stransportMatrixTransient (\tau) \stransportSources (t-\tau) \diff \tau \eqbreak 
& = \left(\stransportMatrixTransient \ast \stransportSources \right)(t).
\label{eq:transienttransportmatrix}
\end{align}

Phasor field virtual waves \cite{Liu2019phasor,Guillen2020Effect,Liu2020phasor} transform  an NLOS imaging problem into a virtual LOS one, by creating a virtual imaging system on a visible diffuse relay surface. It leverages the temporal information \iccvfinal{encoded in} $\pfImpulse$ on the visible surface to propagate a virtual wave from a virtual emitter to a virtual sensor, both placed at the relay wall (\fref{fig:phasor_fields_scheme}). Matrix $\pfImpulse$ is captured by sequentially illuminating a set of points $\xp \in \planeP$ on the visible surface, and measuring the response at a set of points $\xc \in \planeC$ on the same surface. 

Phasor fields leverage the linearity \iccvfinal{and time invariance} of $\pfImpulseFun$ to compute the response of the hidden scene at points $\xc$ in a virtual sensor, for any virtual complex-valued emission profile $\phasorxpt$ as
\begin{align}
\phasorxct = \int_\planeP [\phasorxpt \ast \pfImpulseFun] \diff \xp.
\label{eq:prop_RSD_xp}
\end{align}
Then, any point $\xv$ in the hidden scene can be imaged by propagating the field $\phasorxct$ with an imaging operator $\Phi(\cdot)$ as
\begin{align}
\iccvfinal{\pfImageof{}{\xv, t} = \Phi\left(\phasorxct\right)}.
\end{align}
This imaging operator models a virtual lens and sensor system, and can be formulated in terms of a Rayleigh-Sommerfeld diffraction (RSD) propagator (please refer to the original work for details~\cite{Liu2019phasor}).
Thus, when propagating monochromatic signals of a single frequency $\pfFreq$ this image formation operator $\pfImagingModel\left(\phasorwxct\right)$ becomes \cite{RadarBook} 
\begin{align}
\pfImagingModel\left(\phasorwxct\right) 
& = \left|\int_\planeC \phasorwxct \frac{\pfThinLensFun{\pfFreq}{\xc,\xv}}{|\xv-\xc|} \diff \xc \right|^2, 
\label{eq:imaging_model_xc}
\end{align}
where $\pfThinLens$ is a complex operator that changes the phase of $\phasor$ at frequency $\pfFreq$.
While $\pfImagingModel$ could represent different imaging operators,  we assume a thin-lens model that perfectly focuses a hidden location $\xv$ into a virtual image plane, so  
\begin{align}
\pfThinLensFun{\pfFreq}{\xc, \xv} = e^{-ik|\xv-\xc|}
\label{eq:thinlens_operator}
\end{align}
where $k=\pfFreq/c$ is the wavenumber, with $c$ the speed of light. 
By combining a focused emission signal (\eref{eq:prop_RSD_xp}) with the imaging operator (\eref{eq:imaging_model_xc}) we can image a location $\xv$ in a hidden scene as in traditional LOS imaging setups, as
\begin{align}
\iccvfinal{\pfImageof{}{\xv,t}} = \left| \int_\planeC \int_\planeP \left[\phasorwxpt\! \ast\! \pfImpulseFun\right] \right. \eqbreak 
\left.\frac{\pfThinLensFun{\pfFreq}{\xc, \xv}}{\norm{\xv-\xc}} \dxp  \dxc \right|^2.
\label{eq:imaging_final}
\end{align}

%
Moreover, since this image formation model is entirely virtual and obtained by computation, the emission profile can be any desired function without \iccvfinal{hardware restrictions}. 

%

\section{The Virtual Light Transport Matrix}
\label{sec:method}

\iccv{
Our goal is to compute the NLOS (virtual) light transport matrix $\svtransportMatrix$ for a \textit{hidden} scene, given the measured impulse response function $\pfImpulse$. 
More formally, the NLOS virtual LTM $\svtransportMatrix$ is a two-dimensional matrix where each component $\svtransportMatrix(\xa, \xb)$ represents reflected light at point $\xb$ when a unit illumination focuses at point $\xa$. 
In practice, \iccvfinal{we choose a coaxial configuration} where $\svtransportMatrix$ has a size of $K_\textbf{v}\times K_\textbf{v}$, \iccvfinal{and} $K_\textbf{v}$ is the number of voxels discretizing the hidden scene (i.e., the number of imaged points in three-dimensions).
To achieve our goal, we leverage the transport operators of the phasor fields framework~\cite{Liu2019phasor}, and create virtual projector-camera pairs on the visible relay surface, focusing at points $\xa$ and $\xb$ respectively. }

\begin{figure}[t]
	\centering
	\begin{subfigure}[b]{0.408\linewidth}
		\def\svgwidth{\textwidth}
		\centering
\begingroup%
  \makeatletter%
  \providecommand\color[2][]{%
    \errmessage{(Inkscape) Color is used for the text in Inkscape, but the package 'color.sty' is not loaded}%
    \renewcommand\color[2][]{}%
  }%
  \providecommand\transparent[1]{%
    \errmessage{(Inkscape) Transparency is used (non-zero) for the text in Inkscape, but the package 'transparent.sty' is not loaded}%
    \renewcommand\transparent[1]{}%
  }%
  \providecommand\rotatebox[2]{#2}%
  \newcommand*\fsize{\dimexpr\f@size pt\relax}%
  \newcommand*\lineheight[1]{\fontsize{\fsize}{#1\fsize}\selectfont}%
  \ifx\svgwidth\undefined%
    \setlength{\unitlength}{183.30662525bp}%
    \ifx\svgscale\undefined%
      \relax%
    \else%
      \setlength{\unitlength}{\unitlength * \real{\svgscale}}%
    \fi%
  \else%
    \setlength{\unitlength}{\svgwidth}%
  \fi%
  \global\let\svgwidth\undefined%
  \global\let\svgscale\undefined%
  \makeatother%
  \begin{picture}(1,0.91188894)%
    \lineheight{1}%
    \setlength\tabcolsep{0pt}%
    \put(0,0){\includegraphics[width=\unitlength,page=1]{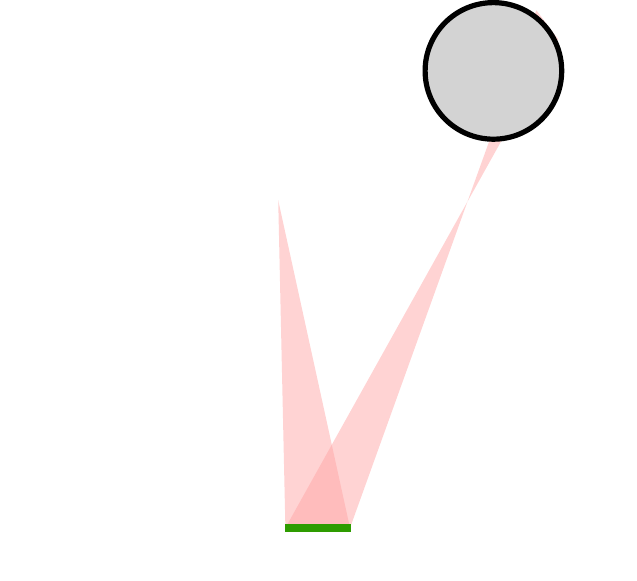}}%
    \put(0.73640167,0.761424){\makebox(0,0)[lt]{\lineheight{1.25}\smash{\begin{tabular}[t]{l}A\end{tabular}}}}%
    \put(0,0){\includegraphics[width=\unitlength,page=2]{narrow_aperture_2D.pdf}}%
    \put(0.40735926,0.65779779){\makebox(0,0)[lt]{\lineheight{1.25}\smash{\begin{tabular}[t]{l}B\end{tabular}}}}%
    \put(0.75258048,0.54303125){\makebox(0,0)[lt]{\lineheight{1.25}\smash{\begin{tabular}[t]{l}$\x_a$\end{tabular}}}}%
    \put(0.47760032,0.54163613){\makebox(0,0)[lt]{\lineheight{1.25}\smash{\begin{tabular}[t]{l}$\x_b$\end{tabular}}}}%
    \put(0.37817699,0.0087797){\makebox(0,0)[lt]{\lineheight{1.25}\smash{\begin{tabular}[t]{l}\footnotesize{Aperture}\end{tabular}}}}%
    \put(0,0){\includegraphics[width=\unitlength,page=3]{narrow_aperture_2D.pdf}}%
    \put(0.01350235,0.53668802){\makebox(0,0)[lt]{\lineheight{1.25}\smash{\begin{tabular}[t]{l}\footnotesize{Focal plane}\end{tabular}}}}%
  \end{picture}%
\endgroup%
		\caption{Narrow aperture (LOS)}
		\label{fig:narrow_aperture_2D}
	\end{subfigure}\hspace{0.5em}
	\begin{subfigure}[b]{0.5\linewidth}
		\def\svgwidth{\textwidth}
		\centering
\begingroup%
  \makeatletter%
  \providecommand\color[2][]{%
    \errmessage{(Inkscape) Color is used for the text in Inkscape, but the package 'color.sty' is not loaded}%
    \renewcommand\color[2][]{}%
  }%
  \providecommand\transparent[1]{%
    \errmessage{(Inkscape) Transparency is used (non-zero) for the text in Inkscape, but the package 'transparent.sty' is not loaded}%
    \renewcommand\transparent[1]{}%
  }%
  \providecommand\rotatebox[2]{#2}%
  \newcommand*\fsize{\dimexpr\f@size pt\relax}%
  \newcommand*\lineheight[1]{\fontsize{\fsize}{#1\fsize}\selectfont}%
  \ifx\svgwidth\undefined%
    \setlength{\unitlength}{226.24493697bp}%
    \ifx\svgscale\undefined%
      \relax%
    \else%
      \setlength{\unitlength}{\unitlength * \real{\svgscale}}%
    \fi%
  \else%
    \setlength{\unitlength}{\svgwidth}%
  \fi%
  \global\let\svgwidth\undefined%
  \global\let\svgscale\undefined%
  \makeatother%
  \begin{picture}(1,0.77033826)%
    \lineheight{1}%
    \setlength\tabcolsep{0pt}%
    \put(0,0){\includegraphics[width=\unitlength,page=1]{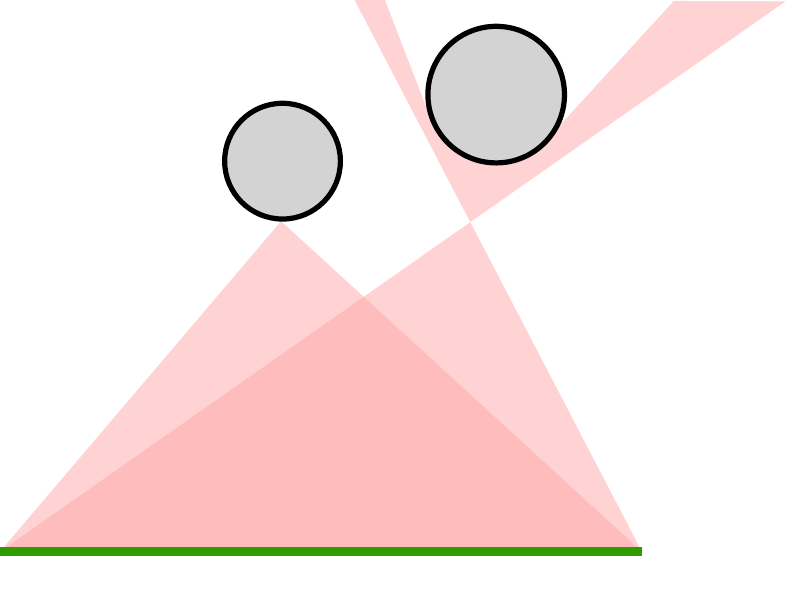}}%
    \put(0.63878627,0.43986251){\makebox(0,0)[lt]{\lineheight{1.25}\smash{\begin{tabular}[t]{l}$\x_a$\end{tabular}}}}%
    \put(0.60184974,0.61989301){\makebox(0,0)[lt]{\lineheight{1.25}\smash{\begin{tabular}[t]{l}A\end{tabular}}}}%
    \put(0.32779659,0.5367624){\makebox(0,0)[lt]{\lineheight{1.25}\smash{\begin{tabular}[t]{l}B\end{tabular}}}}%
    \put(0,0){\includegraphics[width=\unitlength,page=2]{wide_aperture_2D.pdf}}%
    \put(0.40368037,0.44191128){\makebox(0,0)[lt]{\lineheight{1.25}\smash{\begin{tabular}[t]{l}$\x_b$\end{tabular}}}}%
    \put(0.30995395,0.00843317){\makebox(0,0)[lt]{\lineheight{1.25}\smash{\begin{tabular}[t]{l}\footnotesize{Aperture}\end{tabular}}}}%
    \put(0,0){\includegraphics[width=\unitlength,page=3]{wide_aperture_2D.pdf}}%
    \put(0.00495219,0.42782847){\makebox(0,0)[lt]{\lineheight{1.25}\smash{\begin{tabular}[t]{l}\footnotesize{Focal plane}\end{tabular}}}}%
  \end{picture}%
\endgroup%
		\caption{Wide aperture (NLOS)}
		\label{fig:wide_aperture_2D}
	\end{subfigure}
	\caption{(a) LOS projector-camera setups are capable of sharply illuminating and imaging elements of a scene at different depths onto the same focal plane, due to their narrow apertures. (b) The relay wall in NLOS methods behaves as a wide aperture, where focused illumination may spread through large surfaces out of the focal plane (e.g. red area of object A onto pixel at $\xa$); analogously, large areas of out-of-focus objects may contribute to a single imaged pixel. }
\end{figure}

\paragraph{Challenges.} Turning the visible relay wall into a virtual projector-camera pair is not trivial. The two main challenges are: 
\textbf{1)} While LOS imaging setups use small apertures to limit the amount of defocus for both the projector and the camera (\fref{fig:narrow_aperture_2D}), the impulse response $\pfImpulse$ in the hidden scene is captured over relatively large surfaces $\planeP$ and $\planeC$ with respect to the reconstructed scene size (\fref{fig:wide_aperture_2D}). When using $\pfImpulse$ to propagate light through a thin lens model (\eref{eq:thinlens_operator}), $\planeP$ and $\planeC$ correspond to the effective apertures of light and image propagation, respectively. These larger virtual apertures result in significant out-of-focus illumination and geometry during the imaging process, as  \fref{fig:wide_aperture_2D} shows. 
Moreover, \textbf{ 2)} the maximum resolution in the reconstructions is bounded by the capture density of the impulse function $\pfImpulse$ \cite{Liu2020ICCP}. However, increasing this density is a time-consuming process, limited by the particular characteristics of the sensors. A suboptimal resolution also affects the sharpness of the illumination, and thus on the effective resolution of $\svtransportMatrix$.

\iccv{\paragraph{The virtual LTM.} Given the challenges described above, na\"ively computing the NLOS virtual LTM will include a significant component of out-of-focus light $\sdefocusMatrix$, due to the large aperture baseline. This na\"ive $\widetilde{\svtransportMatrix}$ can be expressed as
\begin{equation}
\widetilde{\svtransportMatrix} = \svtransportMatrix + \sdefocusMatrix. 
\end{equation}
We are interested in obtaining $\svtransportMatrix$ without the influence of undesired defocused light. In our derivations we further decompose $\svtransportMatrix$ into its diagonal and off-diagonal elements as $\svtransportMatrix=\svtransportMatrix_\text{d} + \svtransportMatrix_\text{off}$. 
For convenience, we also decompose $\sdefocusMatrix$ into $\sdefocusMatrix_\text{d}$ and $\sdefocusMatrix_\text{off}$, which represent out-of-focus illumination at diagonal and off-diagonal elements, respectively, having

%
\begin{equation}
\widetilde{\svtransportMatrix} = \svtransportMatrix_{\text{d}} + \svtransportMatrix_{\text{off}}+\sdefocusMatrix_\text{d} +\sdefocusMatrix_\text{off}.
\end{equation}
In camera-projector setups $\svtransportMatrix_{\text{d}} = \svtransportMatrix_{\text{d},1}+\svtransportMatrix_{\text{d},\infty}$ is the sum of direct transport $\svtransportMatrix_{\text{d},1}$ plus multiply-scattered retro-reflection and back scattering $\svtransportMatrix_{\text{d},\infty}$, while $\svtransportMatrix_{\text{off}}$ is the sum of all indirect illumination bounces $\svtransportMatrix_{\text{off}} = \sum_{k=2}^\infty \svtransportMatrix_{\text{off},k}$ \cite{OToole2012}.  Note that since we are in NLOS settings, the direct illumination and indirect components in the hidden scene correspond to 3rd and 4th+ bounce illumination, respectively. }

\paragraph{Modeling the illumination function.}
\label{sec:modeling}
\begin{figure*}[t]
	\begin{subfigure}[b]{0.24\linewidth}
		\def\svgwidth{\textwidth}
		\centering
		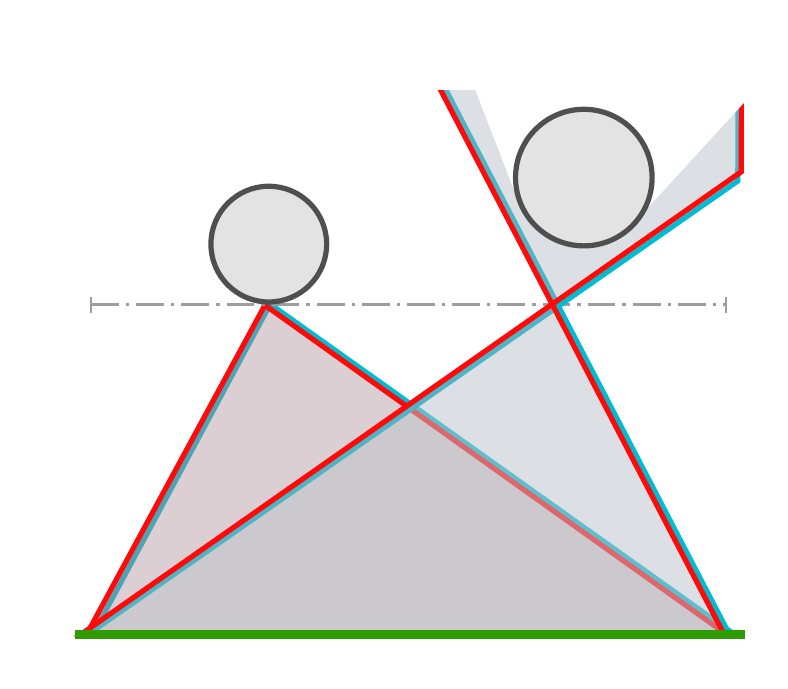%
		\caption{Direct}
		\label{fig:components_direct}
	\end{subfigure}
	\begin{subfigure}[b]{0.48\linewidth}
		\def\svgwidth{\textwidth}
		\centering
		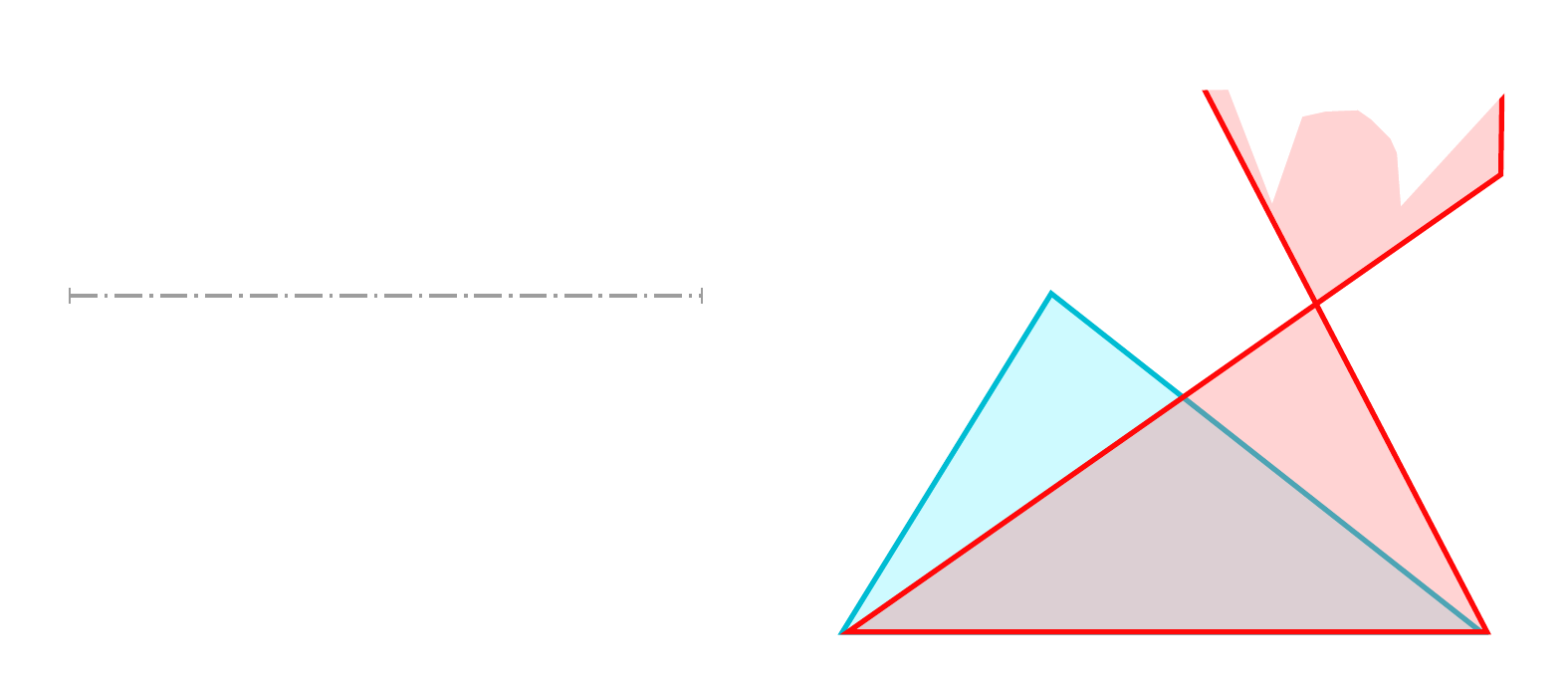%
		\caption{Indirect}
		\label{fig:components_indirect}
	\end{subfigure}
	\begin{subfigure}[b]{0.24\linewidth}
		\def\svgwidth{\textwidth}
		\centering
		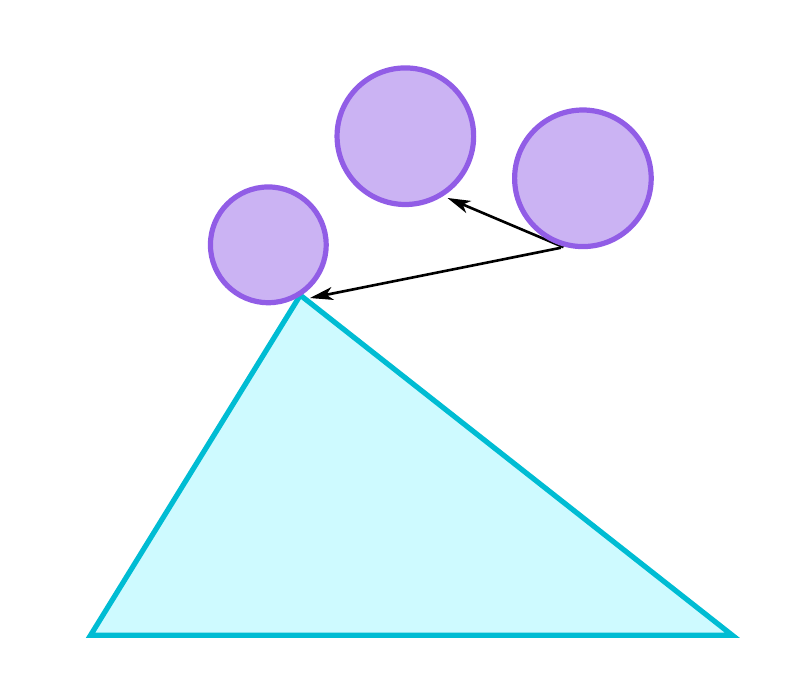%
		\caption{In-focus indirect}
		\label{fig:components_infocus}
	\end{subfigure}
	\caption{
		We transform transient measurements $\pfImpulseFun$ at a visible relay wall (green) into virtual projectors (red) and cameras (blue) facing a hidden scene. We disentangle individual illumination components by computing elements of the virtual transport matrix $\svtransportMatrix(\xa, \xb)$: 
		\textbf{(a)} We compute direct illumination (diagonal of $\stransportMatrix$) by illuminating and imaging the same location $\xv$ ($\xa\equiv\xb$); we gate light paths with the time of flight of $\pathseq{\xp,\xv,\xc}$ to avoid defocusing effects from empty voxels $\xv^\prime$. 
		\textbf{(b)} Left: We compute indirect illumination (off-diagonal of $\stransportMatrix$) by illuminating and imaging different voxels $\xa \neq \xb$, and gating illumination with the time of flight of $\pathseq{\xp,\xa,\xb,\xc}$. Right: Defocusing may still arise at the imaged voxel $\xb$ from paths with similar time of flight $\pathseq{\xp,\xa^\prime,\xb,\xc}$. 
		\textbf{(c)} We isolate in-focus indirect light by using direct illumination as an oracle of the hidden scene geometry (purple), and avoid imaging and illuminating empty locations.} 
	\label{fig:components_schematic}
\end{figure*}


In order to remove the undesirable contribution of unfocused light encoded in $\sdefocusMatrix_\text{d}$ and $\sdefocusMatrix_\text{off}$,  
we derive different virtual illumination functions $\phasorwf{\xp,t}$ \iccv{that enable the computation of diagonal and off-diagonal elements $\svtransportMatrix_{\text{d}}$ and $\svtransportMatrix_{\text{off}}$
}
%
\iccv{Figs. \ref{fig:components_direct} and \ref{fig:components_indirect} illustrate this: (a) diagonal elements, where the virtual projector (red) and camera (blue) simultaneously illuminate and image the same point (voxel) of the hidden scene; (b) off-diagonal elements, where projector and camera focus on different points.}

%
To implement the resulting virtual projector-camera system we extend the parameter space of $\phasorwf{\xp,t}$ to take into account different locations of the voxelized hidden space. In its general form, we then have
\begin{align}
\phasorwf{\xp,t,\dots} = \gatingf{\xp,t,\dots} \frac{\pfThinLensFun{\pfFreq}{\xp,\dots}}{\distf{\xp,\dots}}.
\label{eq:generic_gating_thin}
\end{align}
%
The term $\gating$ represents a real-valued time-dependent gating function,
while $\pfThinLensFun{\pfFreq}{\xp,\dots}$ is a complex-valued thin-lens operator (\eref{eq:thinlens_operator}) which propagates light from points in the illumination aperture $\xp \in \planeP$ to specific locations of the hidden space at a distance $\dist$. 
%
In the following, we develop the illumination functions $\phasorwf{\xp,t,\dots}$ required  \iccv{to compute the specific elements of the virtual LTM of NLOS scenes}.

\subsection{Diagonal elements}
\label{sec:direct_separation}
%

In our NLOS setting, \iccv{we compute the diagonal elements $\stransportMatrix_{\text{d}}$ (i.e. $\stransportMatrix(\xv,\xv)$)} by focusing both the virtual projector and the virtual lens at the same position $\xv$ (see \fref{fig:components_direct}). 
As discussed before (and shown in \fref{fig:wide_aperture_2D}), this can introduce significant out-of-focus illumination $\sdefocusMatrix_\text{d}$ \iccv{(see $\xv^\prime$ in \fref{fig:components_direct})}. To avoid such contribution, we use a gating function centered at the time-of-flight of the path $\pathseq{\xp,\xv,\xc}$, thus removing the contribution of longer path lengths. From an NLOS perspective, this gating function isolates three-bounce light paths between the laser and the SPAD. Since the rise of backprojection algorithms \cite{Arellano2017NLOS,Velten2012nc,Buttafava2015,Laurenzis2014area}, isolating the third-bounce illumination (also known as temporal focusing) has been a recurrent approach to improve robustness in NLOS reconstructions~\cite{Pediredla2019SNLOS,Liu2019phasor}. Implementations of this temporal focusing include explicitly selecting temporal bins from captured transients in backprojection methods \cite{Velten2012nc}, gating the illumination during the acquisition process \cite{Pediredla2019SNLOS}, or convolving virtual propagators with Gaussian pulses \cite{Liu2019phasor}. We follow this last option, and use unit-amplitude Gaussian gating functions $\gatingf{t^\prime, t}$ centered at $t$ with standard deviation $\sigma$ \iccvfinal{so approximately 99\% of the Gaussian covers four times the propagation wavelength.}
%
%
The resulting illumination function $\phasorwf{\pathseq{\xp,\xv,\xc},t}$ thus becomes
\begin{align}
\phasorwf{\pathseq{\xp,\xv,\xc},t} =\gatingf{\tof_d, t} \frac{\pfThinLensFun{\pfFreq}{\xp, \xv}}{\norm{\xp-\xv}},
\label{eq:direct_illum_profile}
\end{align}
with $\tof_d = \left(\norm{\xc-\xv} + \norm{\xp-\xv}\right)/c$.

Finally, combining the imaging process in \eref{eq:imaging_final} with the illumination function in \eref{eq:direct_illum_profile} we compute each diagonal element of $\svtransportMatrix_{\text{d}}$ as
\begin{align}
\iccvfinal{\pfImageof{d}{\xv, t\!:=\!0}} = \left| \int_\planeC \int_\planeP \right. & \left[\phasorwf{\pathseq{\xp,\xv,\xc},t} \! \ast\! \pfImpulseFun\right] \eqbreak  
& \left. \frac{\pfThinLensFun{\pfFreq}{\xc, \xv}}{\norm{\xv-\xc}} \dxp \dxc \right|^2.
\label{eq:direct_light}
\end{align}
\iccv{\paragraph{Discussion:} In practice, this links our virtual NLOS LTM with the confocal camera proposed by Liu et al.~\cite{Liu2019phasor}.
Note that \eref{eq:direct_light} actually computes the direct illumination $\svtransportMatrix_{\text{d},1}$ component of $\svtransportMatrix_{\text{d}}$, while removing the contribution of $\svtransportMatrix_{\text{d},\infty}$.
Computing $\svtransportMatrix_{\text{d},1}$ is required for geometry estimation from the third bounce. Our {objective}, however, is to also separate the indirect component to enable NLOS light transport analysis.
}

\subsection{Off-diagonal elements}
\label{sec:indirect_separation}

\iccv{The off-diagonal matrix $\svtransportMatrix_{\text{off}}$ models all light reflected off $\xb$ when focusing light on another point $\xa$, i.e. $\svtransportMatrix(\xa,\xb)$ for $\xa\neq\xb$. It represents the indirect illumination of the scene~\cite{OToole2012}, and in our NLOS configuration corresponds to light paths of the form $\pathseq{\xp,\xa,...,\xb,\xc}$.} 
\iccv{As in the case of the diagonal elements $\svtransportMatrix_{\text{d}}$, direct computation of the off-diagonal elements might result in significant out-of-focus contribution $\sdefocusMatrix_\text{off}$. To minimize this, we design a second illumination function as follows. 

First, we decompose $\svtransportMatrix_{\text{off} }= \svtransportMatrix_{\text{off},2} + \svtransportMatrix_{\text{off},3-\infty}$, where $\svtransportMatrix_{\text{off},2}$ represents the contribution of two-bounce illumination, and $\svtransportMatrix_{\text{off},3-\infty}$ the remaining higher-order bounces.}
We focus on two-bounce illumination with 4-vertex paths of the form $\pathseq{\xp,\xa,\xb,\xc}$ (\fref{fig:components_indirect}, left), and discard the contribution of higher-order scattering since it decreases exponentially with each bounce. 

An illuminated point $\xa$ will bounce light towards a point $\xb$ following a sub-path $\vvec_{ab}=\xb-\xa$ with time of flight $\tof_{ab} = |\xb-\xa| c^{-1}$. 
\iccv{To compute the coefficient $\svtransportMatrix_{\text{off},2}(\xa,\xb)$ of our NLOS LTM} we need to \textbf{1)} focus direct illumination on the source point $\xa$, \textbf{2)} gate light paths centered at the time of flight of the 4-vertex path $\pathseq{\xp,\xa,\xb,\xc}$, and \textbf{3)} image point $\xb$. Since the positions of all four vertices are known, paths of this form can be isolated by using a narrow temporal gating function as (we remove the path-dependence for clarity) 
\begin{equation*}
\tof_{i,4} = \left(|\xa-\xp| + |\xb-\xa| + |\xs-\xb|\right) c^{-1}.
\end{equation*}
%
For gating, we use the same unit-amplitude Gaussian gating function as in the diagonal elements, but centered at $\tof_{i,4}$ as $\gatingf{\tof_{i,4}, t}$. A gating function for higher-order bounces is described in \sref{sec:gating_highbounces}.
%
The illumination and imaging operators therefore correspond to thin-lens propagators $\pfThinLensFun{\pfFreq}{\xp, \xa}$ and $\pfThinLensFun{\pfFreq}{\xc, \xb}$, respectively. This yields the following illumination function
\begin{align}
\phasorwf{\pathseq{\xp,\xa,\xb,\xc},t} =\gatingf{\tof_{i,4}, t} \frac{\pfThinLensFun{\pfFreq}{\xp, \xa}}{\norm{\xp-\xa}}.
\label{eq:phasor_indirect_short}
\end{align}
\iccv{Finally, we estimate the off-diagonal coefficients $\svtransportMatrix_{\text{off},2}(\xa,\xb) \approx \pfImageof{\xa}{\xb}$ as}
\begin{align}
\pfImageof{\xa}{\xb} = \left| \int_\planeC \int_\planeP \right. & \left[\phasorwf{\pathseq{\xp,\xa,\xb,\xc},t}\! \ast\! \pfImpulseFun\right] \eqbreak
 & \left. \frac{\pfThinLensFun{\pfFreq}{\xc, \xb}}{\norm{\xs-\xb}} \dxp  \dxc \right|^2.
 \label{eq:indirect_single_source}
\end{align}
\iccv{This represents the indirect contribution at $\xb$ from direct illumination at $\xa$ after a single light bounce from $\xa$ to $\xb$}. To obtain the total indirect illumination of a single point $\xa$ from the entire scene, we need to compute the corresponding column of $\svtransportMatrix_{\text{off}}$ for all imaged points $\xb$ (yellow column in \fref{fig:components_indirect}, left). 
%


\paragraph{Discussion.} While this method separates illumination with a time-of-flight $\tof_{i,4}$, it is not necessarily restricted to the path $\pathseq{\xp,\xa,\xb,\xc}$. As \fref{fig:components_indirect} (right) shows, when focusing illumination at an \textit{empty} location $\xa$, no light is reflected towards $\xb$ from $\xa$. 
%
%
Instead, given the large illumination apertures $\planeP$, light may be significantly defocused over larger areas past $\xa$  (marked in red on object A). This in turn may lead to light paths $\pathseq{\xp, \x^\prime_a, \xb, \xc}$  with the same time of flight $\tof_{i,4}$ contributing to the imaged location $\xb$ (large imaging apertures $\planeC$ may cause a similar problem).
In the following we show how to mitigate these effects by leveraging our direct light estimations (\sref{sec:direct_separation}) to isolate in-focus indirect illumination (i.e. surface to surface).

\subsection{In-focus off-diagonal elements}
\label{sec:guided_indirect_separation}

While our gating procedure removes most defocused off-diagonal components, there may still be non-negligible illumination in $\sdefocusMatrix_\text{off}$ produced by focusing on empty regions of the scene. 
Direct-only illumination (\sref{sec:direct_separation}) allows us to obtain an estimation $\pfImage_d$ (\eref{eq:direct_light}) of in-focus geometry in our voxelized space $\ROI$. 
Since most empty regions have only a very small contribution in $\pfImage_d(\xv)$ (as a result of capture noise), we can use $\pfImage_d(\xv)$ as an oracle of the subspace $\nonEmpty$ of non-empty voxels as 
\begin{align}
\nonEmpty = \{\xv \in \ROI \vert \pfImage_d(\xv) > \varepsilon\},
\end{align}
where $\varepsilon$ represents a very small value. 

To constrain propagation in the transport matrix $\svtransportMatrix_{\text{off},2}$ to in-focus indirect illumination, we use this subspace $\nonEmpty$ to construct a 1D binary mask $\maskNonEmpty$ of size $K_\textbf{v}$, which we apply to columns and rows of $\svtransportMatrix_{\text{off},2}$ (\fref{fig:components_infocus}). The resulting 2D mask corresponds to the outer product
\begin{align}
\mathbf{M}_{i} = \maskNonEmpty \otimes \maskNonEmpty.
\end{align}
We can therefore compute the indirect contribution of all visible geometry using the same functions as in \sref{sec:indirect_separation}, limited to the subset of all non-empty locations $\xb \in \nonEmpty$ (\fref{fig:components_infocus}, purple), by traversing such non-empty locations $\xa \in \nonEmpty$ and accumulating their contribution (\eref{eq:indirect_single_source}) as
\begin{align}
\pfImageof{i}{\xb} = \sum_{\xa \in \nonEmpty} \pfImageof{\xa}{\xb}.
\end{align}

This allows to accurately reconstruct \iccv{in-focus two-bounce indirect illumination $\svtransportMatrix_{\text{off},2}$} in a hidden scene even in the presence of large virtual apertures, since we explicitly avoid focusing light and camera on empty voxels.



\section{Results}
\label{sec:results}

In the following we demonstrate the performance of our framework in both simulated and real scenes under different scanning configurations of the impulse function $\pfImpulse$. All the results shown here have been obtained using the illumination and gating functions introduced in \sref{sec:direct_separation} to \ref{sec:guided_indirect_separation}. 

\iccvfinal{We compute the simulated scenes using a publicly available transient renderer} \cite{Jarabo2014}. Unless stated otherwise, we compute the impulse response $\pfImpulse$ over a uniform $32\times32$-grid of SPAD positions and a $32\times32$-grid of laser positions; in both cases, they cover areas between $1\!\times\!1$ to $2\!\times\!2$ meters on the relay wall. This 2D-by-2D topology enables focusing both image and illumination, thus fully exploiting the principles of our framework. 
%

\iccv{\fref{fig:teaser} shows a complex hidden scene depicting a crowded living room. Probing the diagonal of $\svtransportMatrix$ we compute the direct illumination component  $\svtransportMatrix_{\text{d}}$, which allows us to recover the bodies at different depths (top), despite occlusions and cluttering. We then isolate indirect light transport when illuminating different points in the scene (A, B and C) by probing their specific columns $\svtransportMatrix_{\text{off}}$ (middle). 
Last, masking out off-diagonal elements at different distances from the diagonal~\cite{OToole2012}, we isolate indirect light transport 
within different path length intervals (bottom).
}

\begin{figure}[t]
	\centering
	\includegraphics[width=0.9\linewidth,page=2]{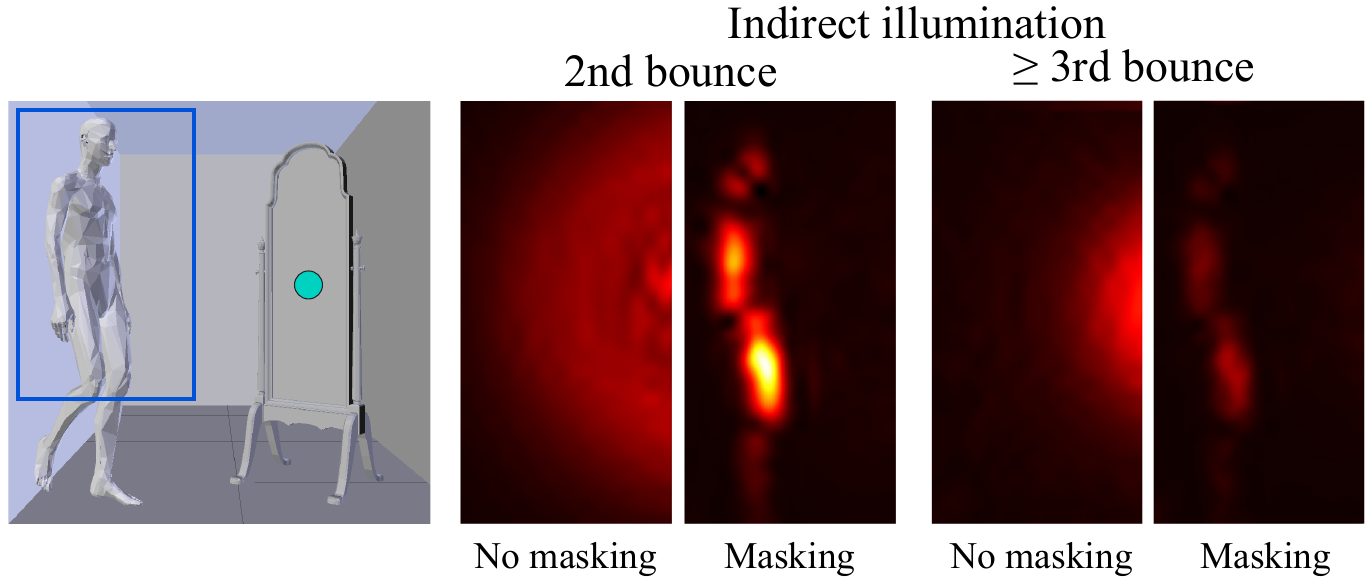}
	\caption{\iccv{
We illustrate effects of isolating indirect illumination at surfaces by illuminating the center of a flat surface that reflects light towards a mannequin. Our masking operation using direct light as an oracle of geometry locations removes most of the out-of-focus illumination in the scene, both in first-order (2nd-bounces) and higher-order (3 or more bounces) indirect illumination.
	}}
	\label{fig:reflector_ablation}
\end{figure}
\begin{figure}[t]
	\centering
	\includegraphics[width=0.7\linewidth]{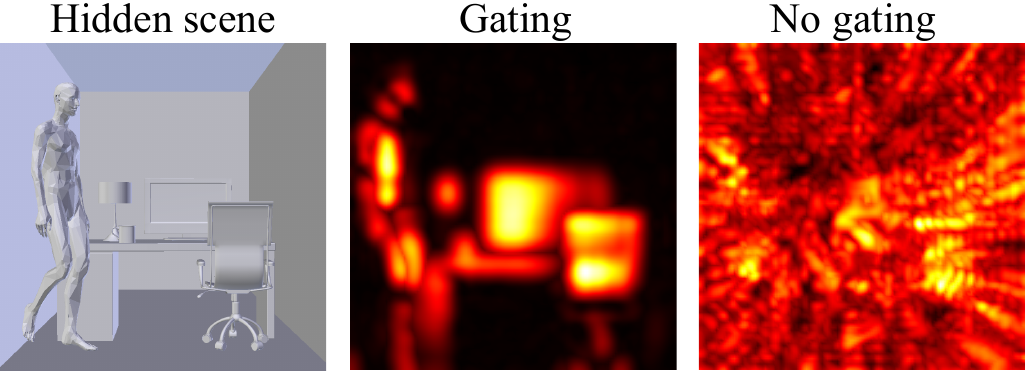}
	\caption{\iccvfinal{Our gating functions (center) are fundamental to mitigate out-of-focus effects produced by large aperture conditions (right).}
	}
	\label{fig:direct_LOS_conditions}
\end{figure}

\iccv{\fref{fig:reflector_ablation} shows how our masking procedure (\sref{sec:guided_indirect_separation}) removes the contribution of defocus components $\sdefocusMatrix_\text{off}$, for 2nd-bounce and higher-order indirect illumination. We illuminate the center of a planar reflector to scatter indirect light towards a mannequin. By using direct light as an oracle of geometry locations we remove most out-of-focus illumination in our indirect light estimations.}

\begin{figure}[t]
	\centering
	\includegraphics[width=0.95\linewidth]{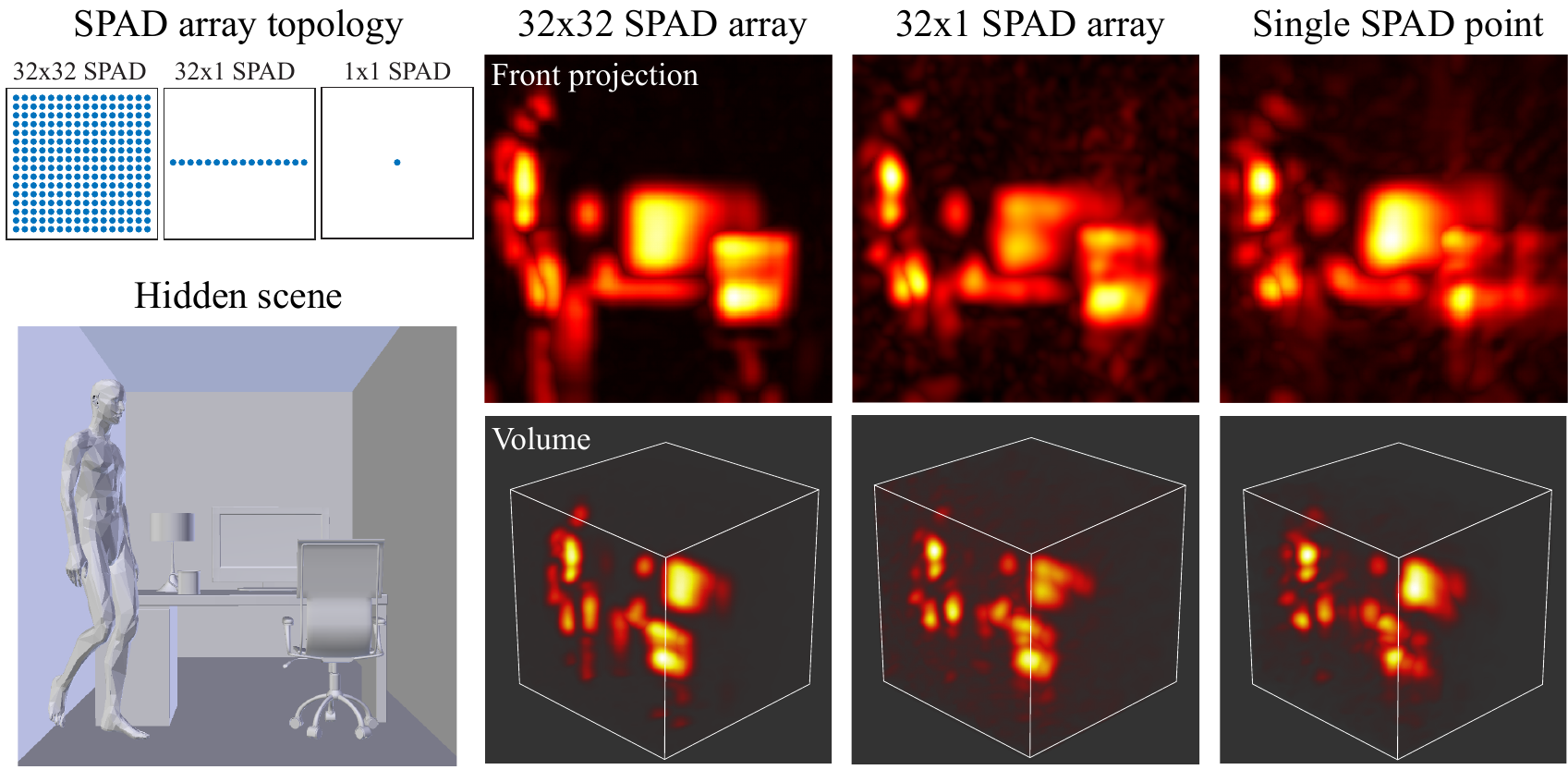}
	\caption{\iccv{Examples with simpler capture setups consisting of 1D arrays and a single SPAD. The resulting LTMs lead to slightly degraded but consistent results.}}
	\label{fig:office}
\end{figure}

\begin{figure}[t]
	\centering
	\includegraphics[width=0.85\columnwidth,page=1]{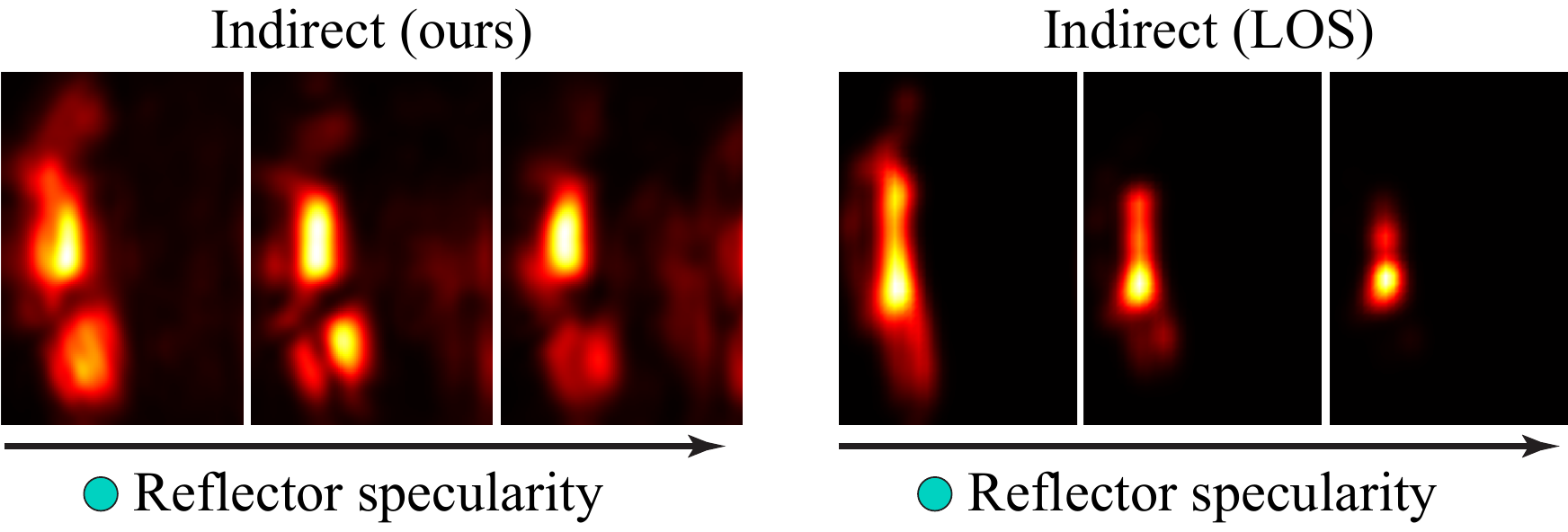}
	\caption{\iccv{Comparison of our NLOS framework against a reference LOS simulation  observing the scene in \fref{fig:reflector_ablation}, left, varying the specularity of the reflector. Our method is consistent with the LOS reference data. }}
	\label{fig:validation}
\end{figure}

\iccv{
Our framework can also be used with simpler capture setups, such as 1D arrays or even single SPADs (\fref{fig:office}). In the case of 1D SPAD arrays the illumination focuses on straight lines perpendicular to the orientation of the array. The resulting LTMs lead to slightly degraded but consistent separation of illumination. }

\vspace{-0.25em}
\paragraph{Validation:}
\iccv{In \fref{fig:validation} we evaluate qualitatively our off-diagonal estimation $\svtransportMatrix_{\text{off}}$, including the removal of the defocus component $\sdefocusMatrix_\text{off}$, by comparing against a LOS render of the hidden scene (\fref{fig:reflector_ablation}, left) for increasing reflector specularity. Our NLOS estimation (left) shows comparable results with the LOS ground truth (right). Note that validating virtual (phasor field) LOS against actual (optical) LOS setups is challenging, since some scene features fall into the null measurement space and thus cannot be recovered~\cite{Liu2019CVPR}.}

\vspace{-0.25em}
\paragraph{Real captured scenes:}
\begin{figure}[t]
	\centering
	\includegraphics[width=\columnwidth]{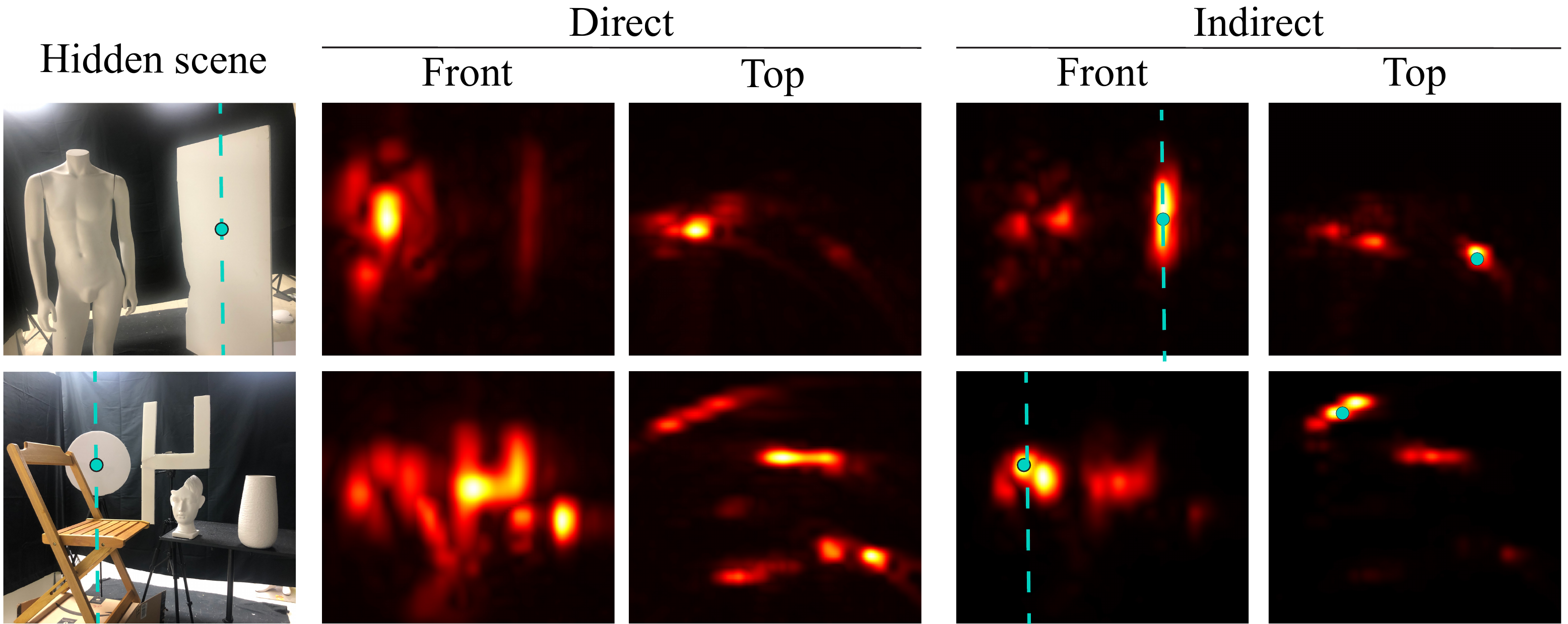}
	\caption{Two real scenes, captured with a 2D laser grid and a 1D SPAD array. \iccv{Blue marks indicate the coordinate where light is focused for indirect illumination computation.} Our framework allows to extract the direct component and isolate the indirect transport from specific points in the scene, even from distant objects.}
	\label{fig:real_scenes}
	\vspace{-0.5em}
\end{figure}
In our real captures our illumination source consists of a OneFive Katana HP pulsed laser operating at 532 nm and 700 \si{mW}, scanning a $24 \times 24$ grid over a $1.9 \times 1.9$ meters area on the relay wall. We use a 1D SPAD array at the wall center, with 28 pixels horizontally spanning 15 \si{cm}. 
A PicoQuant HydraHarp 400 Time-Correlated Single Photon Counting (TCSPC) records the signal coming from the SPADs. 
The effective temporal resolution of our system is 85 ps. The total exposure time was set to 50 seconds per scene. 
\fref{fig:real_scenes} shows the results for two different scenes. Using a horizontal 1D SPAD array focuses illumination over a \textit{vertical} line (dotted line in the figure) instead of a point. Despite this operational limitation, our method is capable of sharply imaging both the direct and indirect components of light transport in complex hidden scenes, including indirect transport from distant objects (such as the circular panel in the second scene).


\section{Conclusions}
\label{sec:conclusions}

We have introduced a framework to obtain the light transport matrix of hidden scenes, coupling the well-established LOS transport matrix with recent NLOS forward propagation methods. This enables direct application of many existing LOS techniques in NLOS imaging setups. 

We overcome the challenges posed by the wide aperture conditions of NLOS configurations by designing virtual imaging systems that mimic traditional narrow-aperture projector-camera setups. 
\iccv{We enable tailoring virtual illumination functions to capture the diagonal and off-diagonal elements of the light transport matrix, roughly representing direct and indirect illumination (first- and higher-order \iccvfinal{bounces}) in the hidden scene, respectively.}
\iccv{In addition, we demonstrate probing techniques for extracting different illumination components of hidden scenes \iccvfinal{featuring} cluttering and lots of occlusions, as well as far/near-range decomposition of indirect light transport.}
Our framework can be directly used for improving NLOS reconstructions, or analyzing the reflectance of hidden surfaces. 


\paragraph{Limitations and future work: } Our work shares resolution limitations similar to other active illumination NLOS methods. Like them, the topology and scanning density of the captured impulse response define the maximum spatial frequency of our virtual projectors and cameras. 
\iccv{While we remove most out-of-focus residual from the off-diagonal components, direct light may lead to overestimation of indirect components due to ambiguous out-of-focus paths falling at geometry locations. 
Also, our current implementation computes the LTM by brute-force sampling each column, resulting in a cost of $O(K_\textbf{v}^2$). Although the computation of $\svtransportMatrix$ can be restricted to diagonals and selected columns, analyzing the properties of the virtual LTM (following e.g., \cite{Wang2009kernel,Peers2009Compressive,o2010optical,OToole2015}) and exploiting line focusing using a 1D SPAD array (see \fref{fig:office}) are interesting avenues of future work for dramatically speeding up computations.}

We have shown that our method works well with currently available 1D SPAD arrays or even single-pixel sensors; the full deployment of 2D SPAD arrays (currently prototypes) will be instrumental in migrating classic LOS imaging systems to NLOS scenarios. 
%
Combining our formulation with the design of new scanning patterns \cite{Liu2020ICCP} and better SPAD models \cite{Hernandez2017SPAD} could lead to optimal light-sensor topologies to improve the estimation of transport matrices in hidden scenes. Last, designing new illumination functions to isolate higher order bounces remains an exciting avenue of future work, which may extend the capabilities of NLOS systems to enable imaging objects around a \emph{second} corner.

\section*{Acknowledgements}
We would like to thank the reviewers for their feedback on the manuscript, as well as Ibón Guillén for discussions at early stages of the project. This work was funded by DARPA (REVEAL HR0011-16-C-0025), the European Research Council (ERC) under the European Union’s Horizon 2020 research and innovation programme (CHAMELEON project, grant agreement No 682080), the Ministry of Science and Innovation of Spain (project PID2019-105004GB-I00), and the University of Wisconsin-Madison Office of the Vice Chancellor for Research and Graduate Education with funding from the Wisconsin Alumni Research Foundation.

\appendix
\section{Gating higher-order bounces}
\label{sec:gating_highbounces}

\iccv{
Here we define a heuristic gating function to compute additional off-diagonal elements corresponding to higher-order bounces  $\svtransportMatrix_{\text{off},3-\infty}$. It selects paths $\pathseq{\xp, \xa, \dots, \xb, \xc}$ with a longer time-of-flight than the 4-vertex paths $\tof_{i,4}$ as
\vspace{-0.25em}
\begin{align}
\gatingf{\tof_{i,3-\infty},t} =
\begin{cases}
0 & t\leq \tof_{i,4} \\
1-\gatingf{\tof_{i,4}, t} & t > \tof_{i,4}
\end{cases},
\label{eq:gating_far}
\end{align}
which is used in \eref{eq:phasor_indirect_short}. Note that since the illuminated and imaged points $\xa$ and $\xb$ remain unchanged (i.e. columns and rows of $\svtransportMatrix_{\text{off},3-\infty}$), the propagation uses the same thin-lens operators $\pfThinLensFun{\pfFreq}{\xp, \xa}$ and $\pfThinLensFun{\pfFreq}{\xc, \xb}$.}
{\small
	\bibliographystyle{ieee_fullname}
	\bibliography{bibliography}
}

\end{document}